\documentclass[pra,preprintnumbers,amsmath,amssymb,showpacs,nofootinbib]{revtex4}
\usepackage{graphicx,bm}

\makeatletter
\def\graphicscale{\twocolumn@sw{0.3}{0.45}}
\makeatother
\begin{document}

\title{Trap-size scaling in confined particle systems at quantum transitions}

\author{Massimo Campostrini and Ettore Vicari}
    \affiliation{Dipartimento di Fisica dell'Universit\`a di Pisa
        and I.N.F.N., Sezione di Pisa,
        Largo Bruno Pontecorvo 2, I-56127 Pisa, Italy}
\date{January 11, 2010}

\begin{abstract}
  We develop a trap-size scaling theory for trapped particle systems at
  quantum transitions.  As a theoretical laboratory, we consider a quantum XY
  chain in an external transverse field acting as a trap for the spinless
  fermions of its quadratic Hamiltonian representation. We discuss trap-size
  scaling at the Mott insulator to superfluid transition in the Bose-Hubbard
  model.  We present exact and accurate numerical results for the XY chain and
  for the low-density Mott transition in the hard-core limit of the
  one-dimensional Bose-Hubbard model. Our results are relevant for systems of
  cold atomic gases in optical lattices.
\end{abstract}


\maketitle


\section{introduction}
\label{intro}

The achievement of Bose-Einstein condensation in diluted atomic vapors of
$^{87}$Rb and $^{23}$Na~\cite{CWK-02} and the impressive progress in the
experimental manipulation of cold atoms in optical lattices (see, e.g.,
Ref.~\cite{BDZ-08} and references therein) have provided a great opportunity
to investigate the interplay between quantum and statistical behaviors in
particle systems.  In these systems, phase transitions are phenomena of great
interest, see, e.g.,
Refs.~\cite{DRBOKS-07,GBMHS-02,PWMMFCSHB-04,KWW-05,HSBBD-06,FWMGB-06,CFFFI-09}.

Phase transitions related to the formation of the Bose-Einstein condensation
in interacting Bose gases at a nonzero temperature, as the one reported in
Ref.~\cite{DRBOKS-07}, are essentially driven by thermal fluctuations, giving
rise to a {\em classical\/} critical behavior, see, e.g., Ref.~\cite{PV-02}.
Quantum fluctuations play a dominant role at $T=0$ transitions, where the
low-energy properties show a {\em quantum\/} critical behavior with a peculiar
interplay between quantum and thermal fluctuations at low $T$, see, e.g.,
Ref.~\cite{Sachdev-book}.  Quantum Mott insulator to superfluid transitions
have been observed in experiments with ultracold atomic gases loaded in
optical
lattices~\cite{GBMHS-02,PWMMFCSHB-04,KWW-05,HSBBD-06,FWMGB-06,CFFFI-09}.

A common feature of the above-mentioned experimental realizations is the
presence of a trapping potential $V(r)$ coupled to the particle density.  For
example, atomic gases loaded in optical lattices are generally described by
the Bose-Hubbard (BH) model~\cite{JBCGZ-98}
\begin{equation}
H_{\rm BH} = -{J\over 2}\sum_{\langle ij\rangle} 
(b_i^\dagger b_j + b_j^\dagger b_i)
    + \sum_i[(\mu + V(r_i))n_i + U n_i(n_i-1)],
\label{bhm}
\end{equation}
where $\langle ij\rangle$ is the set of nearest-neighbor sites and $n_i\equiv
b_i^\dagger b_i$ is the particle density operator.
Far from the origin the potential $V(r)$
diverges, therefore $\langle n_i\rangle$ vanishes and the particles are
trapped.  The inhomogeneity due to the trapping potential strongly
affects the phenomenology of quantum transitions in homogeneous systems.
Correlation functions are not expected to develop a diverging length scale in
the presence of a trap.  Therefore, a theoretical description of how critical
correlations develop in systems subjected to confining potentials is of great
importance for experimental investigations.

We consider the trapping power-law potential
\begin{equation}
V(r) = v^p |r|^p \equiv (|\vec{r}|/l)^p,
\label{potential}
\end{equation}
where $v$ and $p$ are positive constants and $l\equiv 1/v$ is the trap size,
coupled to the particle number.  Experiments are usually set up with
a harmonic potential, i.e., $p=2$.

Let us consider the case in which the system parameters are tuned to values
corresponding to the critical regime of the unconfined system.  In the
presence of a {\em confining potential}, the critical behavior of the {\em
  unconfined} homogeneous system can be observed around the middle of the trap
only in a window where the length scale $\xi$ of the critical modes is much
smaller than the trap size, but sufficiently large to show the universal
scaling behavior.  If $\xi$ is large, but not much smaller than the trap size,
the critical behavior gets somehow distorted by the trap, although it may
still show universal effects controlled by the universality class of the phase
transition of the unconfined system.  In this paper we investigate this issue
within quantum transitions.

In Ref.~\cite{CV-09} we have shown that the critical behavior of trapped
systems at {\em classical\/} continuous transitions can be cast in the form of
a trap-size scaling (TSS), resembling the finite-size scaling theory for
homogeneous systems~\cite{FB-72,GKMD-08}, but characterized by a further
nontrivial {\em trap critical exponent\/} $\theta$, which describes how the
length scale $\xi$ depends on the trap size at criticality, i.e., $\xi\sim
l^\theta$, and which can be estimated using renormalization-group (RG)
arguments.  TSS was supported by numerical results for some lattice gas
models.

In the present paper, we extend the study of the effects of trapping
potentials to {\em quantum\/} critical behaviors at $T=0$ transitions.  We put
forward a TSS scenario to describe how critical correlations develop in large
traps, analogous to the one outlined in Ref.~\cite{CV-09} for {\em classical}
continuous transitions. We then check the validity of the TSS scenario within
quantum transitions by determining the trap-size dependence in a few quantum
models.  As a theoretical laboratory, we consider a quantum XY chain in an
external space-dependent transverse field, acting as a trap for the spinless
fermions of its quadratic Hamiltonian representation.  We present exact and
very accurate numerical results which fully support TSS.  Moreover, we discuss
TSS at the Mott insulator to superfluid transition in the BH model.  In
particular, we present analytical results for the low-density Mott 
transition in the hard-core limit of the one-dimensional BH
model.  The results again support TSS, although the corresponding TSS
functions show peculiar behaviors, like discontinuities in the scaling
particle density, which are related to the quantum nature of the transition.

The paper is organized as follows.  In Sec.~\ref{trapss} we outline the main
ideas of the TSS theory.  In Sec.~\ref{XY} we study the XY chain in a
space-dependent transverse field, which gives rise to an inhomogeneity
analogous to the one arising from a trapping potential in particle systems.
We show how TSS emerges by analytical and very accurate numerical results.  In
Sec.~\ref{BHmodel} we consider the BH model (\ref{bhm}) in a confining
potential. We discuss the general features of the Mott insulator to superfluid
transitions, and present the RG arguments to derive the corresponding trap
exponent $\theta$.  We then present results for the TSS at the low-density
Mott transition in the one-dimensional hard-core BH
model.  Finally, in Sec.~\ref{conclusions} we draw our conclusions.

\section{Trap-size scaling at quantum transitions}
\label{trapss}

Quantum phase transitions arise from a nonanaliticity of the ground-state
energy, where the gap $\Delta=E_1-E_0$ vanishes.  In addition,
continuous quantum transitions have a diverging length scale $\xi$, which
gives rise to peculiar scaling properties, see, e.g., Ref.~\cite{Sachdev-book}.

In a standard general scenario, a quantum $T=0$ transition of a homogeneous
$d$-dimensional system is generally characterized by one relevant parameter
$\mu$, with critical value $\mu_c$ and RG dimension $y_\mu\equiv 1/\nu$, and a
given dynamic critical exponent $z$.  Interesting physical examples are the
Mott insulator to superfluid transitions in BH models.  The quantum critical
behavior is characterized by a vanishing energy scale, for example the gap
$\Delta$, and a diverging length scale $\xi$, behaving respectively as
$\Delta\sim |\bar{\mu}|^{z\nu}$ and $\xi\sim |\bar{\mu}|^{-\nu}$, where
$\bar{\mu}=\mu-\mu_c$, thus $\Delta\sim \xi^{-z}$.

The confining potential gives rise to a space inhomogeneity, thus changing the
scaling behavior of the homogeneous system, which could be only recovered in
the limit of large trap size (keeping fixed the other parameters of the
system).  We want to describe how the critical correlations develop in a
trap. Our starting point is a scaling Ansatz which extends the scaling laws of
homogeneous systems at quantum transitions, to allow for the presence of a
confining potential like (\ref{potential}). We write the scaling law of the
singular part of the free energy density
at the quantum transition as
\begin{equation}
F(\mu,T,v,x) = b^{-(d+z)} F(\bar{\mu} b^{y_\mu},T b^z, v b^{y_v},x/b),
\label{sfreee}
\end{equation}
where $b$ is any positive number, $y_v$ is the RG dimension of the trap
parameter $v$, cf.\ Eq.~(\ref{potential}), and $x$ is the distance from the
middle of the trap. We are neglecting irrelevant scaling fields, because they
do not affect the asymptotic behaviors.  Then, fixing $ v b^{y_v}=1$ and
introducing the trap size $l=v^{-1}$, we obtain the following trap-size
scaling (TSS)
\begin{equation}
F = l^{-\theta (d+z)} {\cal F}(\bar{\mu} l^{\theta/\nu},
Tl^{\theta z},xl^{-\theta}),
\label{freee}
\end{equation}
where $\nu\equiv 1/y_\mu$ and $\theta \equiv 1/y_v$ is the {\em trap
  exponent}.  $\theta$ depends on the universality class of the quantum
transition. It can be computed by evaluating the RG dimensions of the
corresponding perturbation.

Analogously, one can derive the TSS of other observables.  Any
low-energy scale at $T=0$, and in particular the gap, is expected to
behave as
\begin{equation}
\Delta = l^{-\theta z} {\cal D}(\bar{\mu} l^{\theta/\nu}),
\label{Deltasca}
\end{equation}
with ${\cal D}(y)\sim y^{z\nu}$ for $y\to 0$ to match the scaling $\Delta\sim
\bar{\mu}^{z\nu}$ in the absence of the trap.  The correlation length $\xi$
around the middle of the trap, or any generic length scale associated with the
critical modes, behaves as
\begin{eqnarray}
\xi = l^{\theta} {\cal X}(\bar{\mu} l^{\theta/\nu},Tl^{\theta z}),
\label{xisca}
\end{eqnarray}
where ${\cal X}(y,0)\sim y^{-\nu}$ for $y\to 0$.  This implies that at the
$T=0$ quantum critical point the trap induces a finite length scale: $\xi\sim
l^{\theta}$.  In the case of a
generic local operator $O(x)$, with RG dimension $y_o$, we expect that
its expectation value and equal-time correlator behave as
\begin{eqnarray}
&&\langle O(x) \rangle  = l^{-\theta y_o}
{\cal O}(\bar{\mu} l^{\theta/\nu},Tl^{\theta z},xl^{-\theta}),
\label{psca}\\
&&\langle O(x) O(0) \rangle_c
= l^{- 2 \theta y_o} {\cal G}_O(\bar{\mu} l^{\theta/\nu},
Tl^{\theta z},xl^{-\theta}),
\label{gpsca}
\end{eqnarray}
where $x$ measures the distance from the origin, i.e., the middle of the trap.
In the above scaling formulae we have neglected scaling
corrections due to irrelevant perturbations, and possible analytic
contributions.

When $p\to \infty$ the effect of the trapping potential is equivalent to
confine a homogeneous system in a box of size $L=2l$ with open boundary
conditions, whose behavior can be described by a standard finite-size
scaling~\cite{FB-72,PV-02}. Therefore we must have that $\theta\to 1$ when
$p\to \infty$.

The TSS theory outlined in this section provides an effective theoretical
framework to describe quantum critical behaviors in confined systems.
However, it is important to check the validity of its scaling Ansatz, because
the quantum nature of the phenomenon may lead to subtle effects.  For this
purpose, in the following sections, we study the trap-size dependence at some
specific quantum transitions of one-dimensional systems in the presence of a
space-dependent confining potential.

\section{Trap-size scaling in the quantum XY chain}
\label{XY}

The quantum XY chain in a transverse field is a standard theoretical
laboratory for issues related to quantum transitions, see, e.g.,
Ref.~\cite{Sachdev-book}. In this model, an inhomogeneity analogous to
the one arising from a trapping potential in particle systems can be
achieved by considering a space-dependent transverse field, i.e.,
\begin{eqnarray}
H_{\rm XY} = -  \sum_i {1\over 2} [ (1+\gamma) \sigma^x_i \sigma^x_{i+1}
+ (1-\gamma) \sigma^y_i \sigma^y_{i+1}]
- \mu \sigma^z_i  - V(x_i) \sigma^z_i,
\label{Isc}
\end{eqnarray}
where $\sigma^i$ are the Pauli matrices, $0<\gamma\le1$ and $V(x) = v^p |x|^p
\equiv (|x|/l)^p$,  where $v$ and $p$ are positive constants and $l = v^{-1}$
is its length scale.  This model can be mapped into a quadratic Hamiltonian of
spinless fermions by a Jordan-Wigner transformation,
\begin{eqnarray}
&&H =
\sum [ c_{i}^\dagger A_{ij} c_{j}  + {1\over 2}
(c_{i}^\dagger B_{ij} c_{j}^\dagger + {\rm H.c.})], \nonumber \\
&&A_{ij} = 2 \delta_{ij} - \delta_{i+1,j} - \delta_{i,j+1}
+ 2 Q(x_i) \delta_{ij}, \nonumber \\
&&B_{ij} = - \gamma \left( \delta_{i+1,j} - \delta_{i,j+1} \right),
\label{sfi} \\
&&Q(x)= \bar{\mu}+V(x),\quad \bar{\mu}\equiv \mu-1.
\nonumber
\end{eqnarray}
In this picture $\mu$ plays the role of {\em chemical potential} for the
$c$-{\em particles}, and the space-dependent field $V(x)$ acts as a {\em trap}
for the $c$-particles, making their {\em local density} $\langle
n_i\rangle\equiv \langle c^\dagger_i c_i \rangle$ vanish at large distance.
In the following, $l\equiv v^{-1}$ will be considered the {\em trap size}.

In the absence of the trap, the model undergoes a quantum transition at
$\bar{\mu}=0$ in the two-dimensional Ising universality class, separating a
quantum paramagnetic phase for $\bar{\mu}>0$ from a quantum ferromagnetic
phase for $\bar{\mu}<0$.  Around $\bar{\mu}=0$, the quantum critical behavior
shows a diverging length scale, $\xi\sim |\bar{\mu}|^{-\nu}$, and a vanishing
energy scale, $\Delta\sim \xi^{-z}$, where $z$ and $\nu$ are universal
critical exponents: $z=1$ and $\nu\equiv 1/y_\mu=1$ ($y_\mu$ is the RG
dimension of $\mu$).

In the presence of the confining potential, the critical behavior can be
observed only in the limit of large trap size. The trap exponent $\theta$
controlling such limit can be determined by analyzing the RG properties of the
corresponding perturbation at the critical point, which can be represented by
$P_V = \int d^d x\,V(x) \phi(x)^2$, where $\phi(x)$ is the order-parameter
field of the $\phi^4$ theory which describes the behavior of the critical
modes~\cite{CV-09,PV-02}.  Since the RG dimensions of the potential $V(x)$ and
the energy operator $\phi^2$, respectively $y_V=py_v-p$ and
$y_{\phi^2}=d+z-y_\mu$, are related by $y_V + y_{\phi^2} = d+z$, we obtain $p
y_v - p = y_\mu$, and therefore, since $z=1$ and $y_\mu=1$,
\begin{equation}
\theta \equiv 1/y_v =  p/(p+1).
\label{theta}
\end{equation}
Notice that $\theta\to 1$ when $p\to\infty$; this was to be expected, since
when $p\to \infty$ the system becomes equivalent to a homogeneous chain with
$-l\le x \le l$ and open boundary conditions.

We mention that some issues related to the presence of a space-dependent
transverse field in the quantum XY chain model have been addressed in
Refs.~\cite{PKT-07,ZD-08,EIP-09,CKT-09}.

\subsection{The trap-size scaling limit}
\label{tsslim}

The Hamiltonian (\ref{sfi}) can be solved by exact numerical diagonalization,
even in the presence of the trapping potential, following, for example,
Ref.~\cite{LSM-61}.  We look for new canonical fermionic variables 
$\eta_k = g_{ki} c_i^\dagger+ h_{ki} c_i$ which diagonalize the
Hamiltonian, i.e., such that
\begin{eqnarray}
H = \sum_k \omega_k \eta_k^\dagger \eta_k,\qquad \omega_k\ge 0,
\label{diagH}
\end{eqnarray}
and the ground state satisfies $\eta_k |0\rangle =0$ for any $k$.  The
diagonalization is achieved by introducing two sets of orthonormal vectors
$\phi_k$ and $\psi_k$ with components $\phi_{ki}=g_{ki} + h_{ki}$ and
$\psi_{ki}=g_{ki}-h_{ki}$ respectively, satisfying the equations
\begin{equation}
(A+B)\phi_k = \omega_k \psi_k, \qquad
(A-B)\psi_k = \omega_k \phi_k,
\label{abpp}
\end{equation}
where the matrices $A$ and $B$ have been defined in Eq.~(\ref{sfi}).
Thus, we have
\begin{equation}
(A-B)(A+B) \phi_k = \omega_k^2 \phi_k.
\label{phieq}
\end{equation}
Once the vectors $\phi_k$ have been computed, one can easily determine
$\psi_k$ and reconstruct the original $c$-operators as
\begin{equation}
c_i = {1\over 2} \sum_k (\phi_{ki} + \psi_{ki}) \eta_k +
(\phi_{ki} - \psi_{ki}) \eta_k^\dagger,
\label{cieta}
\end{equation}
from which one can compute the correlation functions of the $c$-operators, by
straightforward calculations.

In order to show how TSS emerges, we consider the continuum limit of
Eq.~(\ref{phieq}).  A similar approach was developed in Ref.~\cite{PKT-07} to
study the critical behavior of the quantum XY chain in the presence of a
gradient perturbation.  By rewriting the discrete differences in terms of
derivatives, near the critical point $\bar{\mu}=0$ and for sufficiently small
values of $k$ (this is required by the smoothness hypothesis underlying the
continuum limit), we obtain
\begin{equation}
\left[  4 Q(x)^2 + 4 \gamma \partial_x Q(x)  -
4 \gamma^2 \partial_x^2 - 4 Q(x)\partial_x^2
- 2 \partial_x Q(x) \partial_x - 2 \partial_x^2 Q(x) + ...
\right] \phi_k(x)   =\omega_k^2 \phi_k(x).
\label{schreq}
\end{equation}
This equation has a nontrivial TSS limit: by rescaling
\begin{eqnarray}
&&x = \gamma^{1/(1+p)} l^{p/(1+p)} X, \nonumber\\
&&\bar{\mu} = \gamma^{p/(1+p)} l^{-p/(1+p)} \mu_r, \label{isresc}\\
&&\omega_k = 2 \gamma^{p/(1+p)} l^{-p/(1+p)} \Omega_{k},
\nonumber
\end{eqnarray}
and keeping only the leading terms in the large-$l$ limit (and for small
$\omega_k$), we obtain
\begin{equation}
\left( \mu_r + X^p - \partial_{X} \right)
\left( \mu_r + X^p + \partial_{X} \right) \phi_k(X) = \Omega_{k}^2 \phi_k(X).
\label{trapscaleq}
\end{equation}
Analogously, we obtain the equation for the function $\psi_k(x)$, i.e.,
\begin{equation}
\left( \mu_r + X^p - \partial_X \right)
\psi_k(X) = \Omega_k \phi_k(X).
\label{trapscaleq2}
\end{equation}
It is worth noting that the dependence on $\gamma$ in Eqs.~(\ref{trapscaleq})
and (\ref{trapscaleq2}) disappears, showing the universality of TSS with
respect to changes of the values of $\gamma$ (of course, excluding the value
$\gamma=0$).  One can easily check that the next-to-leading terms in the
large-trap limit give rise to $O(l^{-\theta})$ scaling corrections.
Corrections due to irrelevant perturbations already present in the homogeneous
system are expected to be more suppressed: the exponent associated with the
leading scaling corrections in generic systems belonging to the
two-dimensional Ising universality class is $\omega=2$~\cite{CCCPV-00,PV-02},
thus leading to $O(l^{-2\theta})$ corrections in TSS.

The solution of Eqs.~(\ref{trapscaleq}) and (\ref{trapscaleq2}) allows us to
determine the critical correlations of the original $c$-operators, using
Eq.~(\ref{cieta}).  One can easily determine the large-distance behavior of
the functions $\phi_k(x)$ and $\psi_k(x)$: they decay exponentially as
$\sim\exp(- a |x|^{p+1})$ apart from prefactors.  Their
large-distance decay determines the large-distance behavior of the scaling
correlation functions of the $c$ operators.

A natural question concerns the robustness of the TSS results with respect to
variation of the trapping potential from a simple power law.  For example, let
us consider a potential given by
\begin{equation}
U(x) = (|x|/l)^p + b (|x|/l)^q, \qquad q>p,
\label{vpq}
\end{equation}
where $b>0$ is a constant. A simple analysis based on the rescalings
(\ref{isresc}) shows that the TSS limit is determined by the
smallest power law, while the higher power gives rise to
$O(l^{-(q-p)/(1+p)})$ scaling corrections.

\subsection{Trap-size scaling of observables}
\label{tsso}

In the following we report results obtained by exact diagonalization
at fixed trap size, and compare them with the TSS derived in
Sec.~\ref{trapss}.  The exact numerical diagonalization in the
presence of the trap is done on a chain of length $L$ with open
boundary conditions, with $L$ large enough to have negligible
finite-size effects.

\subsubsection{Trap-size scaling of low-energy scales}
\label{gap}

\begin{figure}[tbhp]
\begin{center}
\leavevmode
\includegraphics*[scale=\graphicscale]{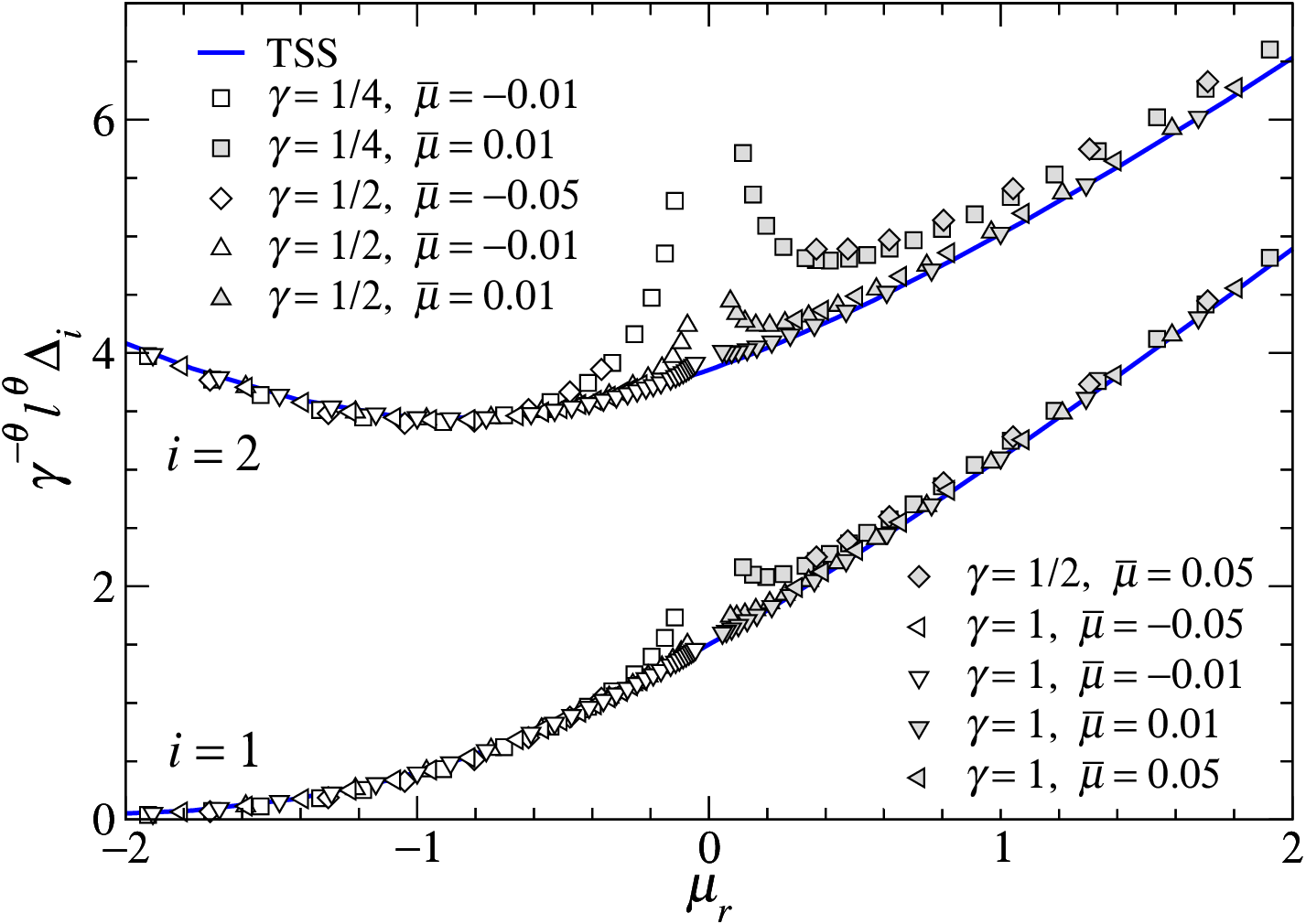}
\hbox to \hsize {\hss}
\includegraphics*[scale=\graphicscale]{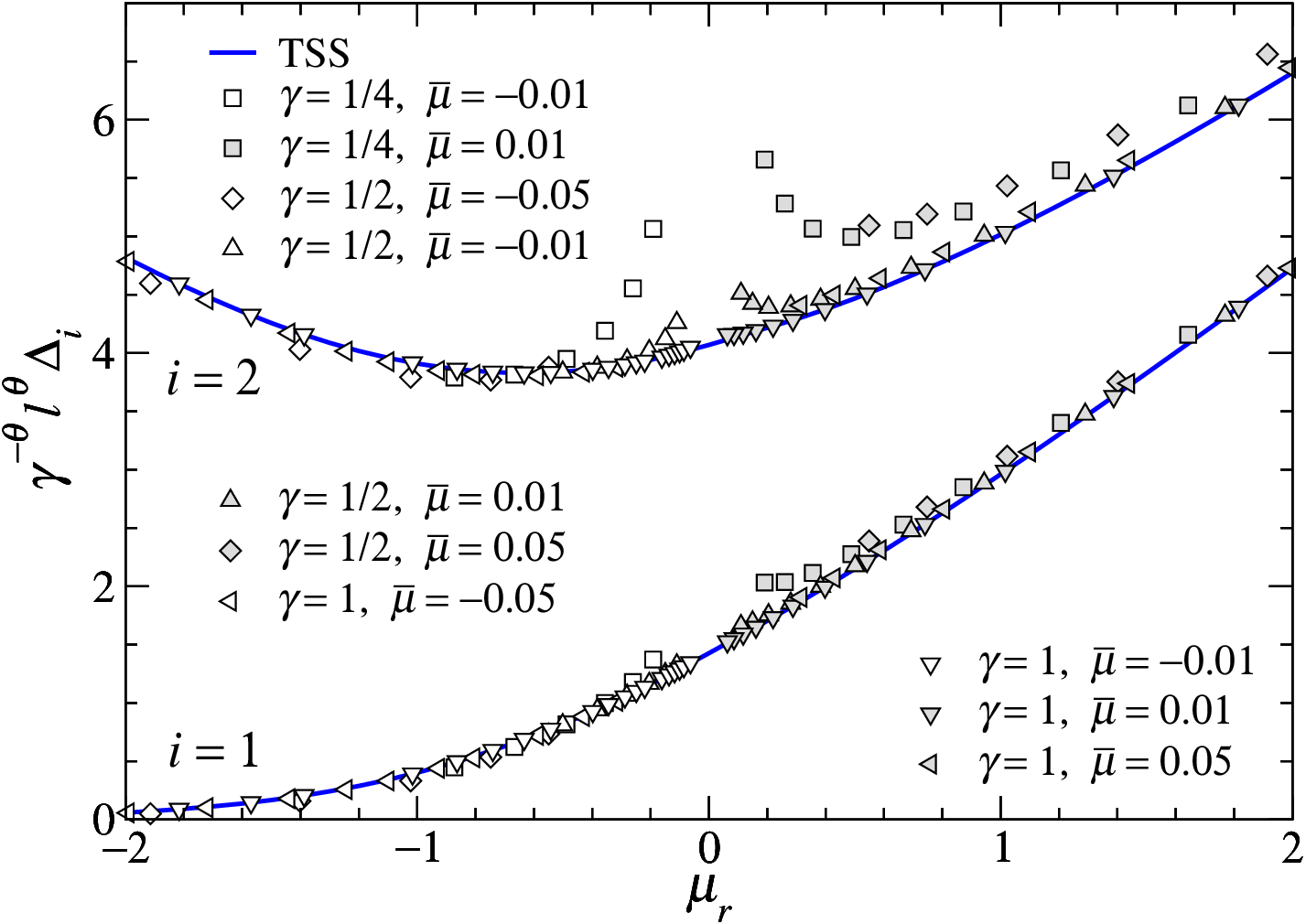}
\caption{(Color online) TSS of the energy differences $\Delta_1\equiv
  E_1-E_0$ and 
  $\Delta_2\equiv E_2-E_0$ for $p=2$ (above) and $p=4$ (below) vs.\ 
  $\mu_r\equiv \gamma^{-\theta} l^{\theta} \bar{\mu}$, for several values of
  $\gamma$ and for $l\ge10$.  Numerical diagonalization results clearly
  approach universal TSS functions in the large-$l$ limit (represented by full
  lines and obtained by extrapolations), with $O(l^{-\theta})$ scaling
  corrections (larger at small $\gamma$ and for higher levels).  }
\label{XYTSS}
\end{center}
\end{figure}

The universal TSS limit obtained after the rescalings (\ref{isresc}) implies
the following asymptotic behavior for any low-energy scale:
\begin{equation}
\Delta \approx \gamma^{\theta} l^{-\theta} {\cal D}(\mu_r),
\quad
\mu_r \equiv \gamma^{-\theta} l^{\theta} \bar{\mu},
\label{deltaisi}
\end{equation}
which is approached with $O(l^{-\theta})$ scaling corrections.  This proves
the scaling behavior (\ref{Deltasca}) obtained by RG arguments.

In Fig.~\ref{XYTSS} we plot the differences among the first few energy levels,
$\Delta_1=E_1-E_0$ and $\Delta_2=E_2-E_0$, obtained by numerical
diagonalization, for trapping potentials with $p=2$ and $p=4$ and for
several values of $\mu$ and $\gamma$.  The numerical results
clearly approaches a universal TSS behavior in the large-$l$ limit, as
predicted by Eq.~(\ref{deltaisi}).  The large-$|\mu_r|$ behavior of the
scaling functions ${\cal D}_i$ related to $\Delta_i$ are ${\cal
  D}_1\sim2\mu_r$ for $\mu_r\to+\infty$, ${\cal D}_1\to0$ for
$\mu_r\to-\infty$, and ${\cal D}_2\sim2|\mu_r|$ for $\mu_r\to\pm\infty$; they
are consistent with the behaviors of the corresponding quantities in the
absence of the trap.  The vanishing of ${\cal D}_1$ for $\mu_r\to -\infty$ is
related to the degeneracy of the ground state in the quantum ferromagnetic
phase.

\subsubsection{Trap-size scaling of the particle density and its correlators}
\label{density}

Other interesting quantities are the particle density and its correlation
function. At the quantum critical point, the particle density is expected to
behave analogously to the energy density in the standard Ising
model~\cite{CV-09}, which presents leading analytic contributions arising from
the analytic part of the free energy.  At $\bar{\mu}=0$ and in the middle of
the trap, we expect
\begin{equation}
\langle n_0 \rangle \equiv  \langle c_0^\dagger c_0 \rangle =
\rho_c(\gamma) + a_n \gamma^{\theta/p} l^{-\theta} + \dots,
\label{nobeh}
\end{equation}
where we used the fact that the RG dimension of the particle density operator
is $y_n=d+z-y_\mu=1$, and the rescalings (\ref{isresc}) to guess the
dependence on $\gamma$ of the amplitude of the scaling term.  $a_n$ is a
constant that does not depend on $\gamma$.  $\rho_c(\gamma)$ is the critical
value of the homogeneous system without trap, for example $\rho_c(1) =
0.18169011...$~\cite{Pfeuty}.  Our numerical results for several values of
$\gamma$ support the scaling behavior (\ref{nobeh}).  Scaling corrections are
(consistent with) integer powers of $l^{-\theta}$.

We have also studied the dependence on the distance from the middle of the
trap, which must be an even analytic function of $x$ at fixed trap size, but
it may present a nontrivial scaling for large $l$.  Analogously to the scaling
at the origin, we expect it to be given by a sum of an analytic behavior and a
nonanalytic term in the trap size, such as
\begin{equation}
\langle n_x - n_0 \rangle  =
\rho_{\rm ns}(x/l) + a_n l^{-\theta}{\cal R}(xl^{-\theta}),
\label{rhoscal}
\end{equation}
where $\rho_{\rm ns}$ is a nonsingular term which we expect to be a
function of $x/l$.  Let us compare this Ansatz with the results from
exact diagonalization.  At small $x$, the latter are consistent with
the expansion
\begin{equation}
\langle n_x - n_0 \rangle \approx b_2(l) x^2 + b_4(l) x^4 + ...,
\label{nxno}
\end{equation}
with
\begin{equation}
b_2(l) = b_{21} l^{-2} \ln l  +  b_{22} l^{-2} + ...,
\qquad b_4(l) = b_{41} l^{-10/3} + ...
\label{nxres2}
\end{equation}
for $p=2$ (dots indicate terms that are more suppressed in the large-$l$ limit
by power laws), and
\begin{equation}
 b_2(l) = b_{21} l^{-12/5} +  b_{22} l^{-16/5} +...,
\qquad b_4(l) = b_{41} l^{-4}\ln l + b_{42} l^{-4} + ...
\label{nxres4}
\end{equation}
for $p=4$. These results should be compared with the expansion of
Eq.~(\ref{rhoscal}) in powers of $x$. They are in substantial agreement; the
only difference is given by the presence of logarithms.  However, this should
not be surprising, because logarithmic corrections to power laws are peculiar
of the two-dimensional Ising universality class, see, e.g., the behavior of
the specific heat, and arise~\cite{Wegner-76} from integer resonances of the
RG dimensions of different perturbations, see also Ref.~\cite{CHPV-02}. Note
that the logarithms appearing in the expansion (\ref{nxno}) show up in the
$x^2$ term for $p=2$ and in the $x^4$ term for $p=4$, where the trap-size
power-law expansions of the two terms in Eq.~(\ref{rhoscal}) present an
accidental degeneracy, just because $3\theta=2$ for $p=2$ and $5\theta=4$ for
$p=4$.  The large-distance behavior of the particle density turns out to be
dominated by the nonuniversal analytic term, indeed at large distance we find
\begin{equation}
\langle n_x \rangle  \approx {\gamma^2 \over 8}
\left(x\over l\right)^{-2p},
\label{ldist}
\end{equation}
while the scaling nonanalytic part is expected to be exponentially suppressed,
as suggested by the large-distance behavior of the solutions of
Eq.~(\ref{trapscaleq}).

\begin{figure}[tbp]
\begin{center}
\leavevmode
\includegraphics*[scale=\graphicscale]{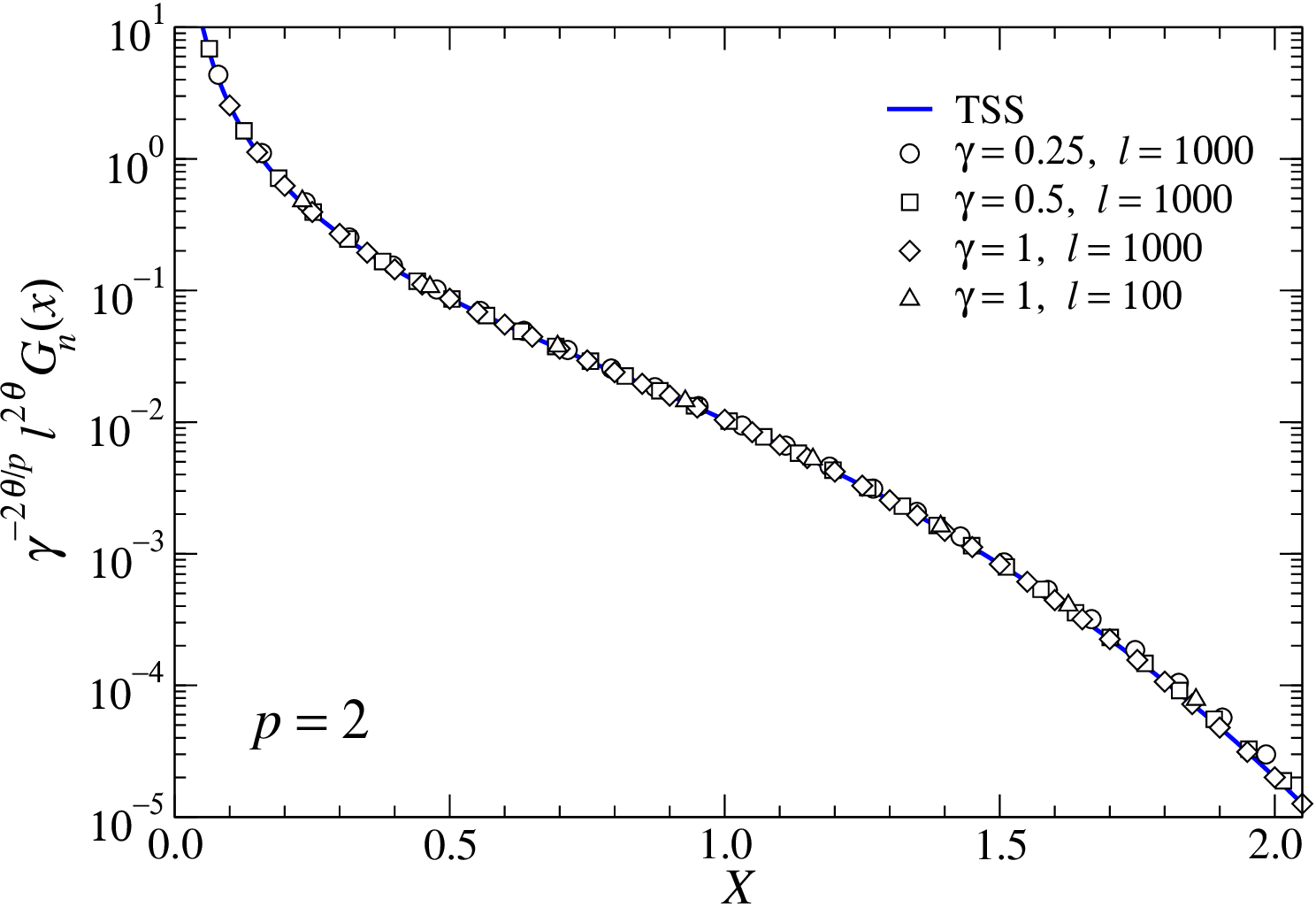}
\hbox to \hsize {\hss}
\includegraphics*[scale=\graphicscale]{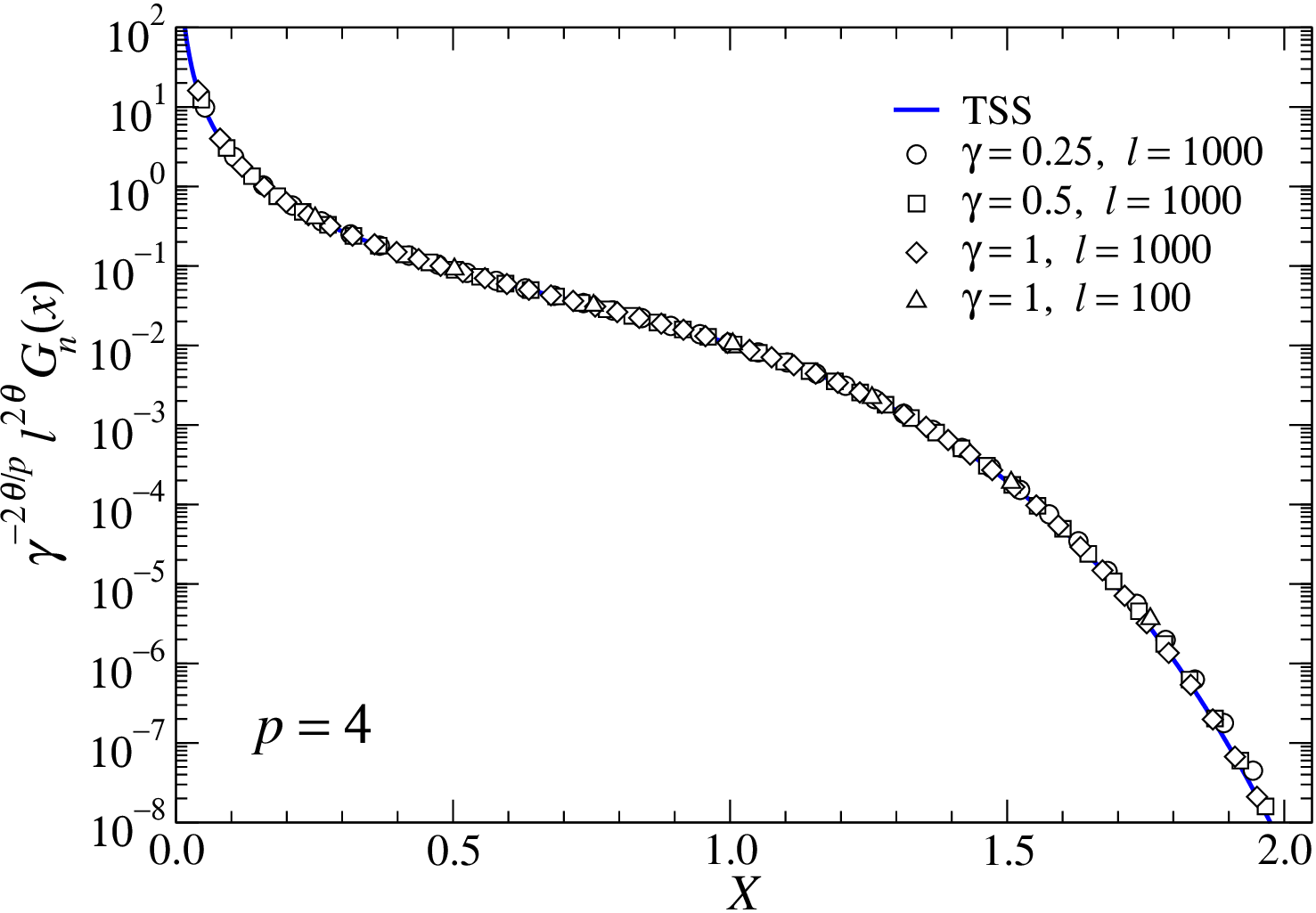}
\caption{(Color online) TSS of $G_n(x)$, cf.\ Eq.~(\ref{nxco}),
 at $\bar{\mu}=0$ vs.\ $X\equiv \gamma^{-\theta/p}
  l^{-\theta} x$, for $p=2$ (above) and
  $p=4$ (below).  Numerical diagonalization results clearly
  approach universal TSS functions in the large-$l$ limit (represented by full
  lines and obtained by extrapolations).  }
\label{TSSGn}
\end{center}
\end{figure}

The static particle-density correlator is not affected by analytic
backgrounds; therefore, according to Eq.~(\ref{gpsca}), we expect that at
$\bar{\mu}=0$ and for $x\ne0$
\begin{equation}
G_n(x) \equiv
    \langle n_0 n_x\rangle - \langle n_0\rangle \langle n_x\rangle
    \approx  \gamma^{2\theta/p} l^{-2\theta}{\cal G}_n(X),
\label{nxco}
\end{equation}
where ${\cal G}_n(X)$ is a universal function.  This is confirmed by the
results of numerical diagonalization, as shown in Fig.~\ref{TSSGn} for
$p=2$ and $p=4$.  At small $X$, ${\cal G}_n(X)\sim 1/X^2$, which is the
behavior in the absence of trap, while at large $X$ it decays very rapidly.

\begin{figure}[tbp]
\begin{center}
\leavevmode
\includegraphics*[scale=\graphicscale]{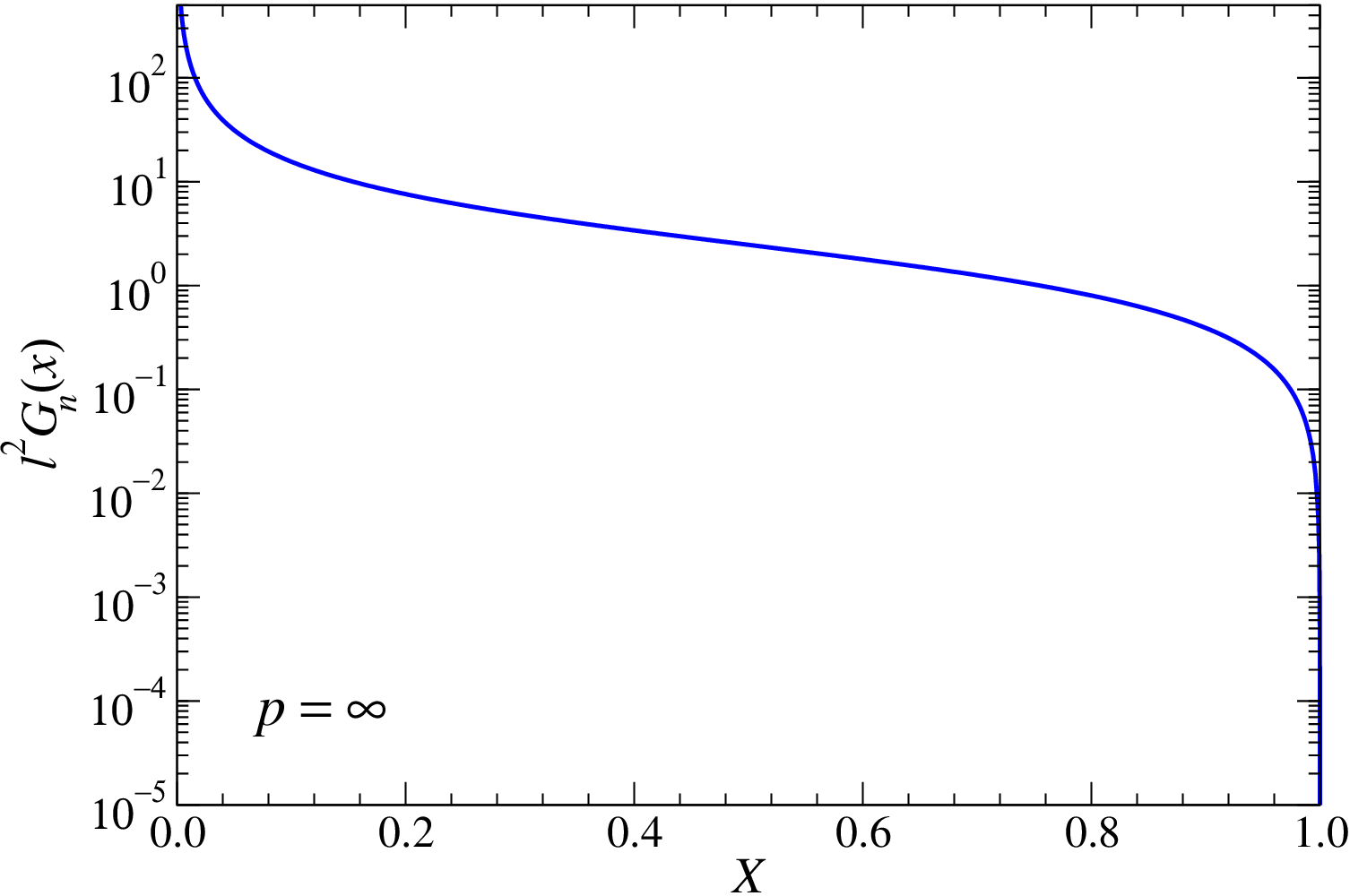}
\caption{(Color online)
The scaling function ${\cal G}_n(X)\equiv l^{2\theta} G_n(x)$
vs.\ $X\equiv x/l^{\theta}$ in the limit $p\to\infty$
where $\theta\to 1$, cf.\ Eq.~(\ref{cfgn}).
}
\label{TSSGninf}
\end{center}
\end{figure}

In the limit $p\to\infty$ the system becomes equivalent to a homogeneous
finite-size chain with $-l\le x\le l$ and open boundary conditions.  Thus, in
this limit we can exploit conformal field theory (CFT), by mapping the
half-plane result for the two-dimensional Ising universality class
\cite{Cardy-84} into the strip, obtaining the correlation
function~\cite{PC-pri}
\begin{equation}
G_n(x) = {\pi^2\over 4l^2} {\cos(\pi X/2) \over \sin^2(\pi X/2)},
\qquad X=x/l.
\label{cfgn}
\end{equation}
Since $\theta\to 1$ for $p\to\infty$, this result agrees with TSS.  Of
course, the corresponding scaling function ${\cal G}_n(X)\equiv l^2G_n(x)$ is
defined for $|X|\le 1$, and vanishes for $X=\pm 1$. It is shown in
Fig.~\ref{TSSGninf}.

\subsubsection{Trap-size scaling of two-point correlation function}
\label{corrfunc}

\begin{figure}[tbp]
\begin{center}
\leavevmode
\includegraphics*[scale=\graphicscale]{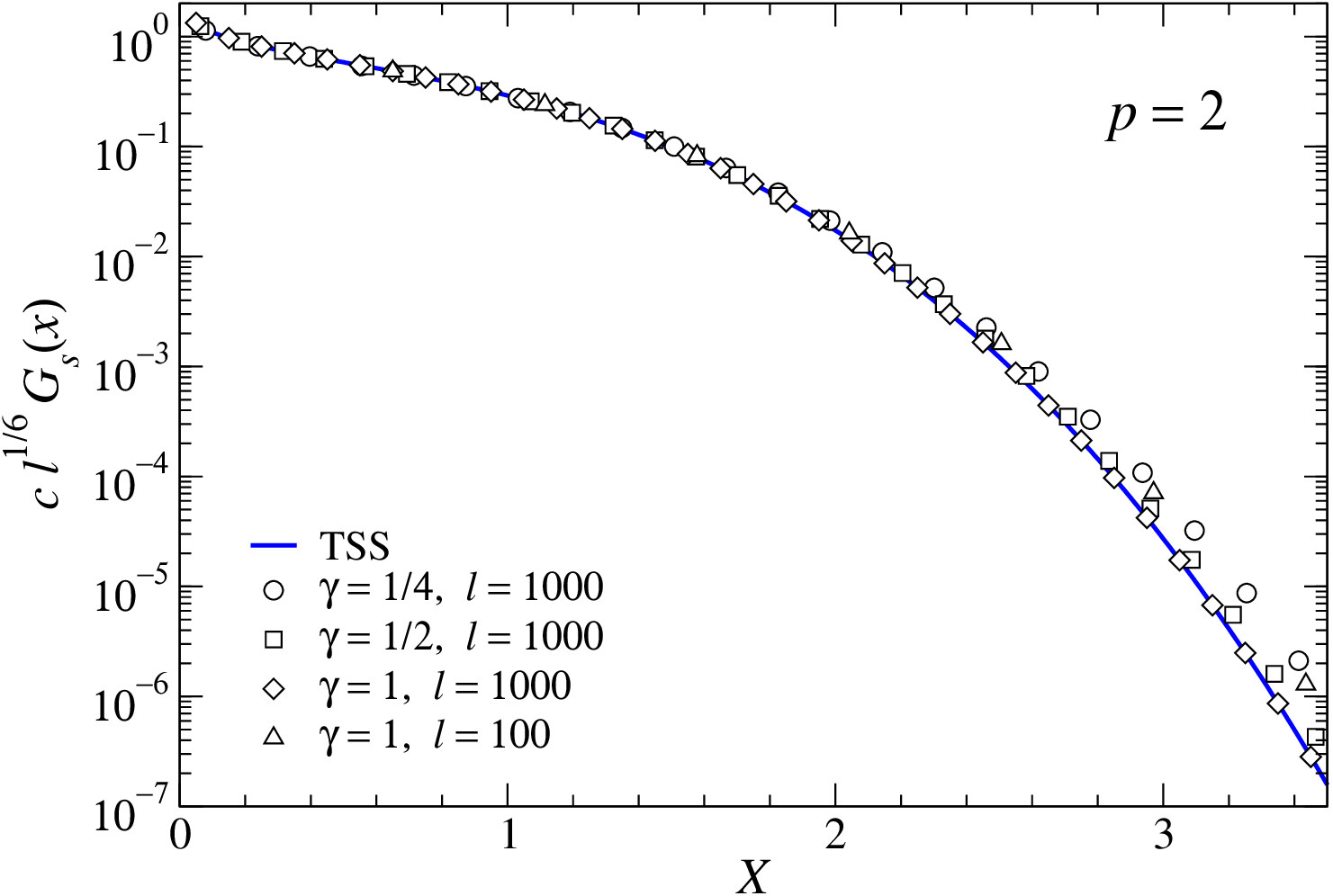}
\hbox to \hsize {\hss}
\includegraphics*[scale=\graphicscale]{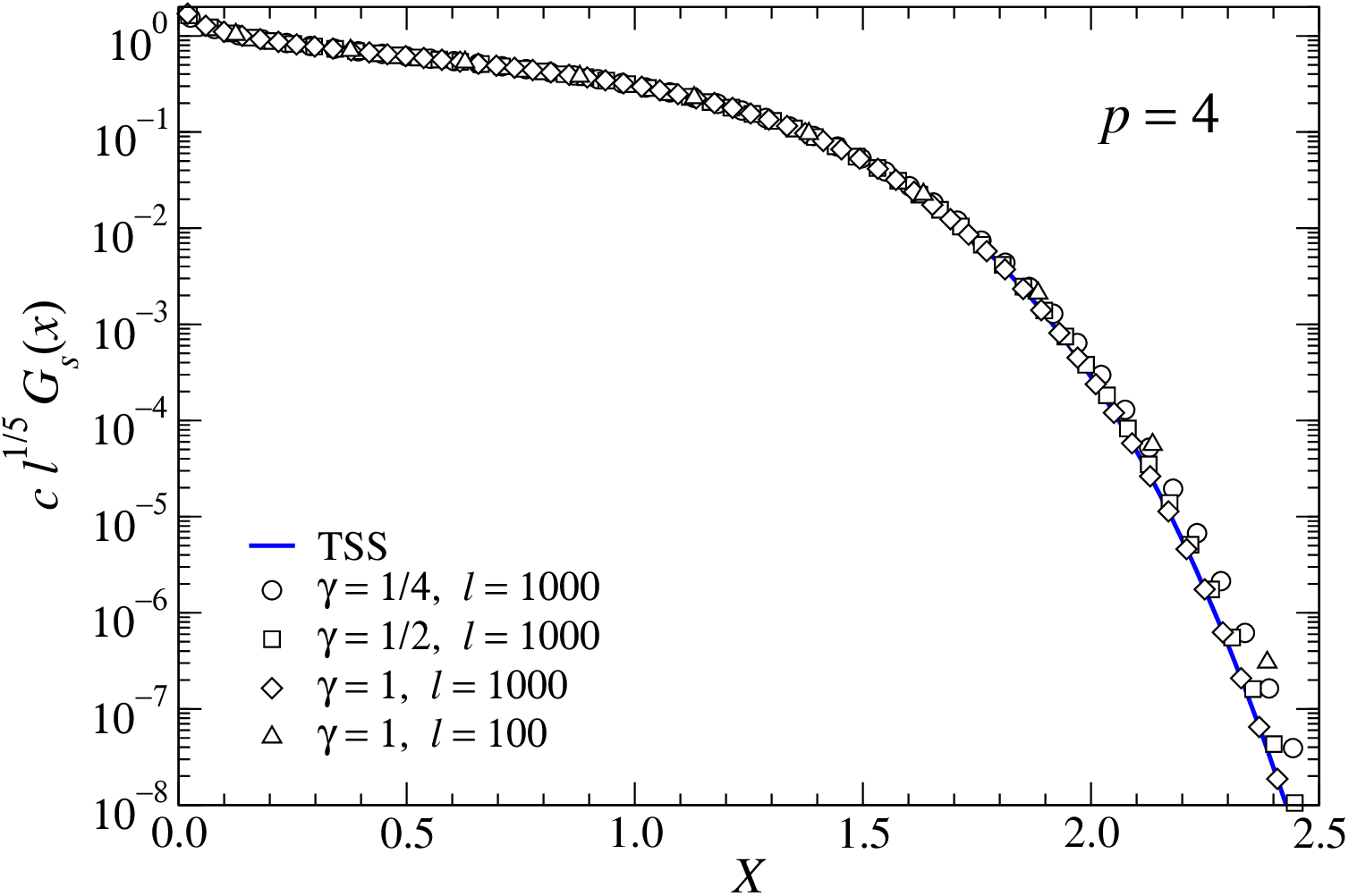}
\caption{(Color online) TSS of the two-point function $G_s(x)$, 
  cf.\ Eq.~(\ref{sxco}), for $p=2$ (above) and $p=4$ (below).
  The universal TSS functions are represented by full lines and
  obtained by extrapolations in the large-$l$ limit.} 
\label{fig:XYcxxscal}
\end{center}
\end{figure}

We have also computed the static correlation function of the spin operator
$\sigma^x$ at $T=0$ and $\mu_c$, following essentially the method of
Ref.~\cite{YR-96}.  According to Eq.~(\ref{gpsca}), $G_s(x)$ 
behaves as
\begin{equation}
G_s(x) \equiv \langle \sigma^x_0 \sigma^x_x \rangle =
a_s l^{-\theta\eta}{\cal G}_s(X),
\label{sxco}
\end{equation}
where we used the fact that the RG dimension of the spin operator $\sigma^x$
is $\eta/2$ with $\eta=1/4$. The constant $a_s$ is not universal, it is
expected to depend on $\gamma$.  Our results from numerical diagonalization
are perfectly consistent with Eq.~(\ref{sxco}): we find that, introducing a
$\gamma$- (and $p$-) dependent normalization $c$, determined
phenomenologically, the combination $c l^{\theta\eta} G_s(x)$ plotted vs.\ 
$X=\gamma^{-1/(1+p)} l^{-\theta}x$, for different values of $x$ (not too
small, e.g., $x>4$), $l$ (large), and $\gamma$, falls on a single curve; see
Fig.\ \ref{fig:XYcxxscal}.
${\cal G}_s(X)$ diverges as $X^{-1/4}$ for
$X\to 0$ (corresponding to the behavior in the absence of the trap) and decay
rapidly at large $X$, as $\exp(-ax^{(p+1)})$, in agreement with the large
distance behavior of the solutions of Eq.~(\ref{trapscaleq}).

One may also consider the susceptibility and second moment correlation length
defined as
\begin{eqnarray}
&&\chi = \int dr \,G_s(r) \approx c_\chi l^{(d-\eta)\theta}+B
=a_\chi l^{3\theta/4}+B,
\label{chis}\\
&&\xi^2 = {1\over 2\chi} \int dr\,r^2\,G_s(r)
\approx a_{\xi^2} l^{2\theta},
\label{xi2}
\end{eqnarray}
where $B$ is a background constant term already present in the homogeneous
system without trap~\cite{PV-02}.
Since the normalization of $G_s(x)$ cancels out, $\xi$ must scale as
$x$, i.e.,
\begin{equation}
\xi = a_\xi \gamma^{\theta/p}  l^{\theta} [ 1 + O(l^{-\theta})],
\label{xissca}
\end{equation}
with $a_\xi$ dependent on $p$ but not on $\gamma$.  The above scaling
behaviors have been verified by numerical computations.

\begin{figure}[tbp]
\begin{center}
\leavevmode
\includegraphics*[scale=\graphicscale]{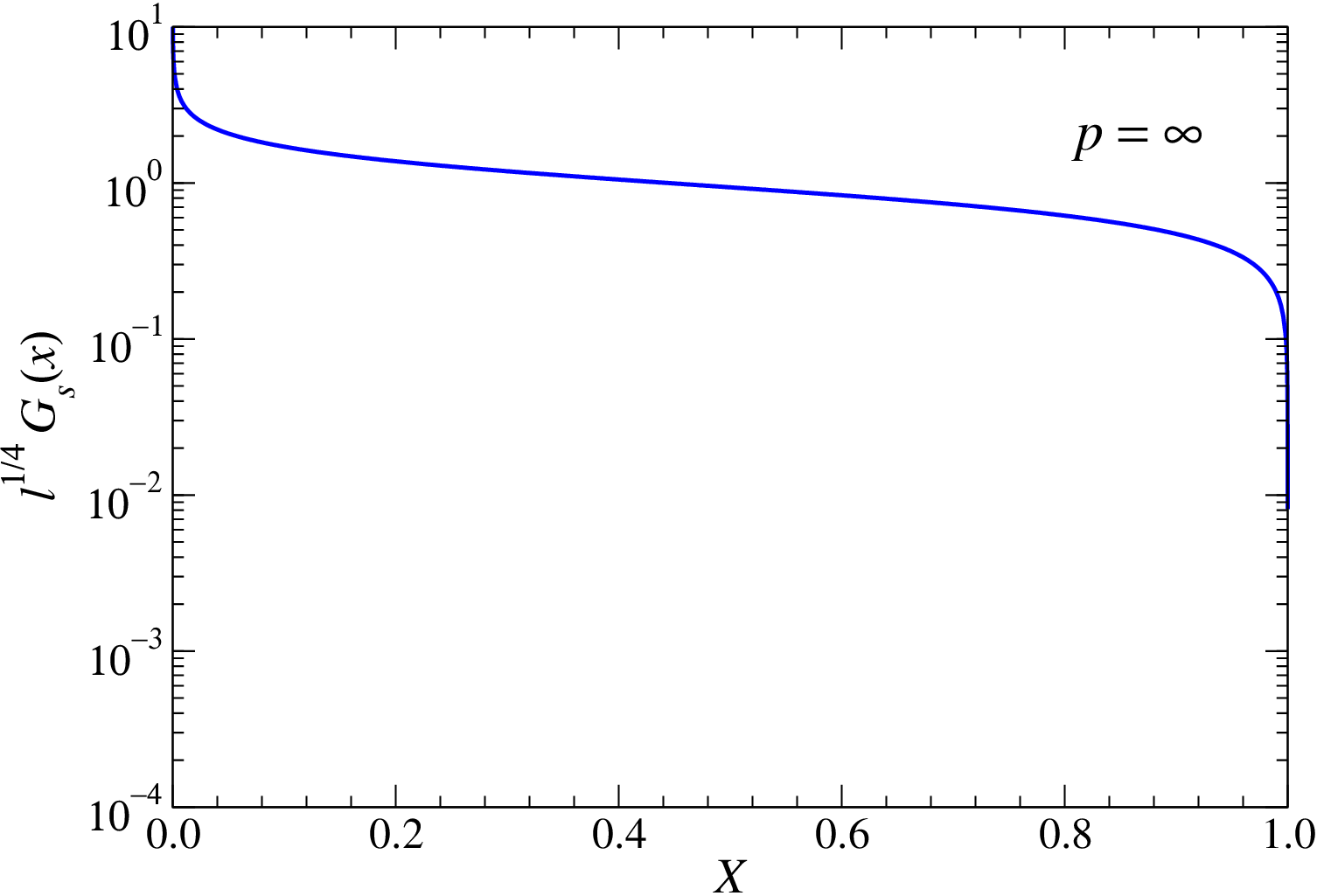}
\caption{(Color online) The scaling function 
   ${\cal G}_s(X)\equiv l^{\theta\eta} G_s(x)$
   vs.\ $X\equiv x/l^{\theta}$ in the limit $p\to\infty$
   where $\theta\to 1$, cf.\ Eq.~(\ref{gspinfty}).}
\label{fig:XYcxxscalinf}
\end{center}
\end{figure}

Finally, the $p=\infty$ limit of the correlation function (\ref{sxco}) can be
computed using CFT, analogously to the density-density correlation function
(\ref{cfgn}), obtaining~\cite{PC-pri}
\begin{eqnarray}
&&G_s(x) =  \left[ {\pi^2\over l^2 16 Y^2 \sin^2 (\pi X/4)}\right]^{1/8}
{\sqrt{1+Y}-\sqrt{1-Y}\over \sqrt{2}},
\nonumber\\
&&Y = \sqrt{1 - \tan^2( \pi X/4)},\qquad X = x/l.
\label{gspinfty}
\end{eqnarray}
The corresponding scaling curve ${\cal G}_s(X) \equiv l^{1/4} G_s(x)$ is
plotted in Fig.~\ref{fig:XYcxxscalinf}.

\subsubsection{Trap-size scaling of quantum entanglement}
\label{entang}

We now discuss quantum entanglement in the presence of the confining
potential.

We divide the chain in two parts of length $l_A$ and $L-l_A$. At $\bar{\mu}=0$,
in the absence of the trap and for open boundary conditions, the entanglement
entropy is given by~\cite{CC-04}
\begin{equation}
S(l_A;L)\approx {c\over 6} \ln \left[L
\sin( \pi l_A/L)\right] + E
\label{ccfo}
\end{equation}
where $c=1/2$ is the central charge.  The constant $E$ depends on $\gamma$:
\begin{equation}
E = {a\over 2} + {1\over 12} \ln (2\gamma/\pi)
\label{bcoe}
\end{equation}
where $a=0.478558...$~\cite{CCD-07}.

In the presence of a trapping potential, and for sufficiently large lattice
sizes $L$, we obtain
\begin{equation}
S(L/2;L)\approx {c\over 6} \ln \xi_e  + E,
\label{ccfot}
\end{equation}
which defines an entanglement length $\xi_e$ at the critical point. Note that
$E$ is the same constant appearing in Eq.~(\ref{bcoe}).  The ratio between
$\xi_e$ and the correlation length $\xi$ defined from the correlation function
of $\sigma^x$, cf.\ Eqs.~(\ref{xi2}) and (\ref{xissca}), should be
universal in the large-$l$ limit, thus independent of $\gamma$.  This
implies that  at criticality $\xi_e$ has the asymptotic behavior
\begin{equation}
\xi_e = a_e \gamma^{\theta/p} l^{\theta} [1+ O(l^{-\theta})],
\label{xie}
\end{equation}
with $a_e$ dependent on $p$ but not on $\gamma$.

The above scenario is fully supported by numerical computations, using the
technique of Refs.\ \cite{EIP-09,Peschel}.

\subsection{Trap-size dependence for $\bm{\mu<1}$}
\label{mumin}

Finally, it is worth discussing the trap-size dependence at values of $\mu<1$,
i.e., $\bar{\mu}<0$, within the quantum ordered phase.  In this case the TSS
theory outlined in Secs.~\ref{trapss} and \ref{tsslim} does not apply, because
the system is not critical in the middle of the trap.  Around the middle of
the trap we have a quantum ferromagnetic phase, up to the point
$\bar{\mu}+V(x)=0$, i.e., up to $|x_c|=(-\bar{\mu})^{1/p} l$, around which
critical fluctuations develop.  Then, as before, at larger distances the
system is in the quantum paramagnetic phase and the particle density vanishes
for $|x|\to \infty$.

\begin{figure}[tbp]
\begin{center}
\leavevmode
\includegraphics*[scale=\graphicscale]{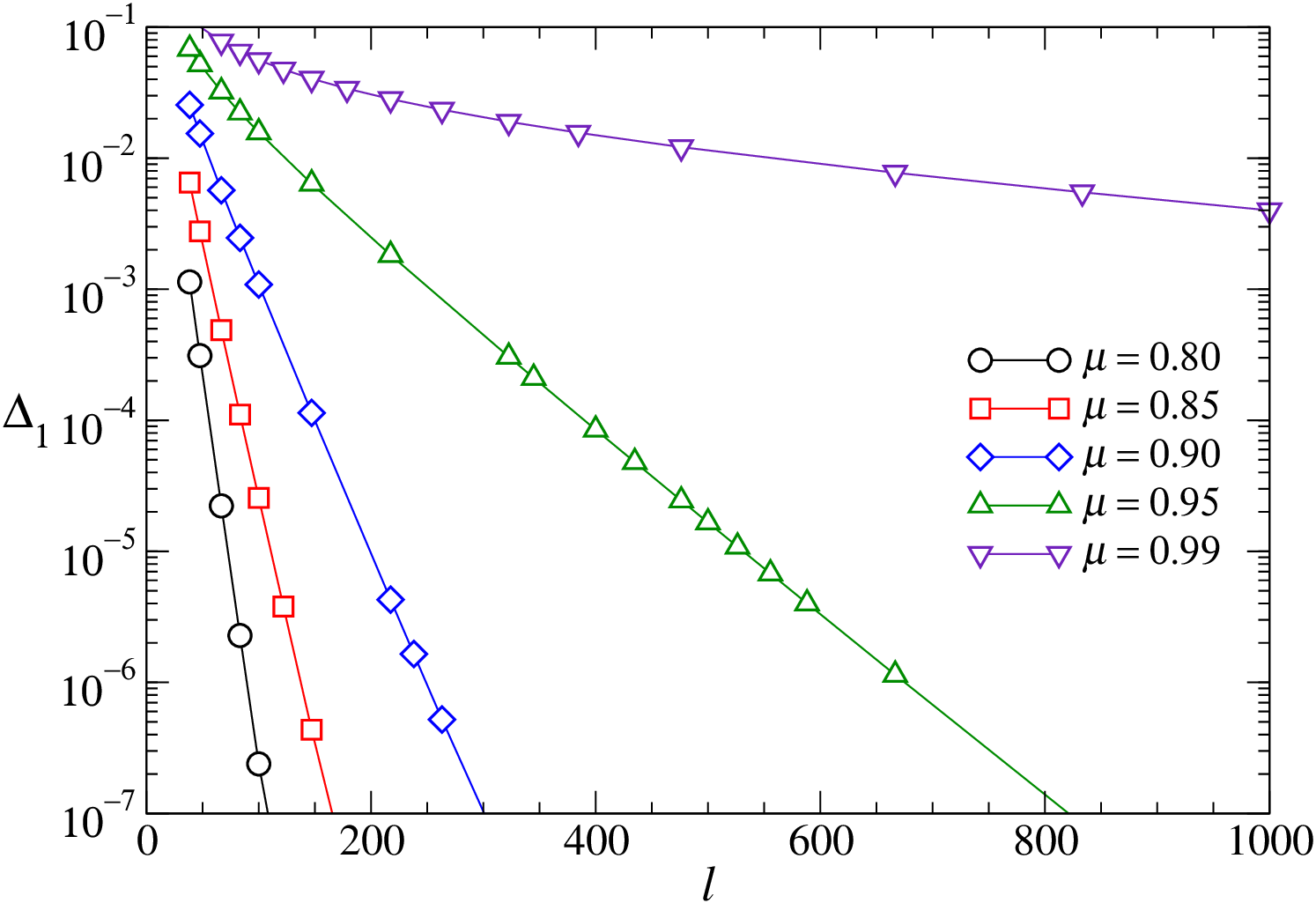}
\hbox to \hsize {\hss}
\includegraphics*[scale=\graphicscale]{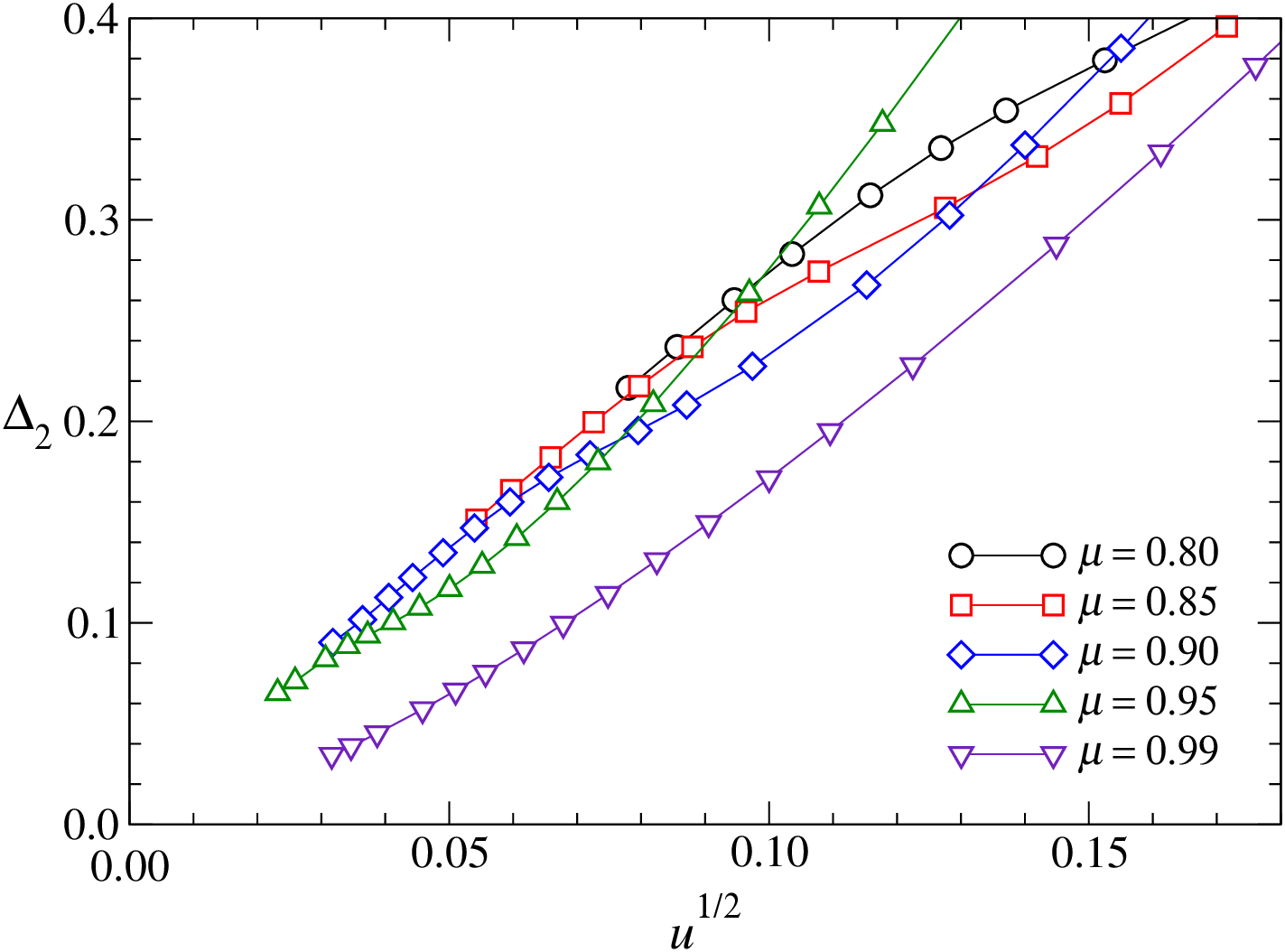}
\vskip1mm
\caption{(Color online)
  Trap-size dependence at $\mu<1$, i.e., in the quantum ferromagnetic phase, of
  the difference of the lowest energy levels for $\mu<1$ and $p=2$:
  $\Delta_1\equiv E_1-E_0$ vs.\ $l$ (above) and $\Delta_2\equiv E_2-E_0$
  vs.\ $u^{1/2}\equiv 2^{1/2}(-\bar{\mu})^{1/4} l^{-1/2}$ (below).
}
\label{deltamin1}
\end{center}
\end{figure}

In Fig.~\ref{deltamin1} we plot the differences of the lowest energy
levels, i.e., $\Delta_1=E_1-E_0$ and $\Delta_2=E_2-E_0$, for a few
values of $\mu<1$ for the harmonic potential, i.e., $p=2$.  $\Delta_1$
vanishes rapidly, apparently as $\Delta_1 \sim \exp(- \kappa l)$,
where the constant $\kappa$ depends on $\mu$.  This behavior is
clearly related to the degeneracy of the ground state in the quantum
ferromagnetic phase in the absence of trapping potential.

It is interesting to note that also $\Delta_2$ appears to vanish in the
large-$l$ limit, although at a much slower rate.  Indeed, our data for $\mu$
not too close to the critical value $\mu=1$ are consistent with the power-law
behavior $\Delta_2\sim l^{-1/2}$.  The vanishing of $\Delta_2$ should be
related to the developing of critical fluctuations at the points $x_c=\pm
(-\bar{\mu})^{1/p} l$ where the sum $\bar{\mu}+V(x)$ vanishes.  Since around
$x_c$ the potential $V(x)=|x|^p/l^p$ can be approximate by a linear potential,
i.e.,
\begin{equation}
V(x)\approx u (x-x_c), \qquad u = p (-\bar{\mu})^{1/p}/l,
\label{vlin}
\end{equation}
the exponent $1/2$ should be related to the RG scaling dimensions of the
parameter $u$ of the linear potential.  Using RG arguments analogous to those
leading to Eq.~(\ref{theta}), we obtain $y_u=2$, see also Ref.~\cite{PKT-07}.
This explanation is further supported by the fact that the data for $\Delta_2$
for different values of $\mu<1$ appear to merge to a unique curve as a
function of $u$ for sufficiently large trap size $l$, as shown in
Fig.~\ref{deltamin1} ($\mu=0.99$ clearly displays crossover effects because it
is very close to the critical value $\mu=1$).

\section{Trap-size scaling in the Bose-Hubbard model}
\label{BHmodel}

\subsection{General features}
\label{gf}

We now discuss TSS within the Bose-Hubbard (BH) model (\ref{bhm}) at the Mott
insulator to superfluid transitions. A review of results for the BH model can
be found in Ref.~\cite{Sachdev-book}.  In the homogeneous BH model without
trap, the low-energy properties of the transitions driven by the chemical
potential $\mu$ are described by a {\em nonrelativistic\/} U(1)-symmetric
bosonic field theory~\cite{FWGF-89}, whose partition function is given by
\begin{eqnarray}
&&Z = \int [D\phi] \exp \left( - \int_0^{1/T} dt \, d^dx
\,{\cal L}_c \right) ,
\\
&&{\cal L}_c = \phi^* \partial_t \phi + {1\over 2 m}
|\nabla \phi|^2 + r |\phi|^2  + u |\phi|^4,
\label{lb}
\end{eqnarray}
where $r \sim \mu-\mu_c$.  The upper critical dimension of this bosonic theory
is $d=2$. Thus its critical behavior is mean field for $d>2$.  For $d=2$ the
field theory is essentially free (apart from logarithmic corrections), thus
the dynamic critical exponent is $z=2$ and the RG dimension of the
coupling $\mu$ is $y_\mu = 2$.  In $d=1$ the theory turns out to be equivalent
to a free field theory of nonrelativistic spinless fermions, from which one
infers the RG exponents $z=2$ and $y_\mu=2$.

The special transitions at fixed integer density (i.e., fixed $\mu$) belong to
a different universality class, described by a {\em relativistic}
U(1)-symmetric bosonic field theory~\cite{FWGF-89},
whose Lagrangian is
\begin{equation}
{\cal L}_c = |\partial_t \phi|^2 +  v^2
|\nabla \phi|^2 + r |\phi|^2  + u |\phi|^4,
\label{lbo2}
\end{equation}
for which $z=1$ and $y_\mu = 1/\nu_{XY}$ where $\nu_{XY}$ is the correlation
length exponent of the $D=d+1$ XY universality class. Thus $\nu_{XY}=1/2$ for
$d=3$ (i.e., mean-field behavior apart from logarithms), $\nu_{XY}=0.6717(1)$
in the $d=2$ case~\cite{CHPV-06}, and formally $\nu=\infty$ for the
Kosterlitz-Thouless transition~\cite{KT-73} at $d=1$.

In the presence of a confining potential, theoretical and experimental results
have shown the coexistence of Mott insulator and superfluid regions when
varying the total occupancy of the lattice, see, e.g.,
Refs.~\cite{JBCGZ-98,BRSRMDT-02,WATB-04,RM-04,GKTWB-06,FWMGB-06}.  However, at
fixed trap size, the system does not develop a critical behavior with
diverging length scale~\cite{BRSRMDT-02,WATB-04}. Criticality is recovered
only in the limit of large trap size, where the critical behavior can be
described in the framework of the TSS theory.  The trap exponent $\theta$ can
be determined by an analysis of the corresponding RG perturbation, i.e.,
\begin{equation}
\int d^dx\,dt\,V(x) |\phi(x)|^2,
\label{rgpert}
\end{equation}
leading to the relation
\begin{equation}
p y_v - p =  d+z-y_{|\phi|^2} = y_\mu,
\label{yU}
\end{equation}
where $y_\mu$ is the RG dimension of the coupling $r$ of the quadratic term of
Lagrangian (\ref{lb}) or (\ref{lbo2}). Therefore,
\begin{equation}
\theta\equiv {1\over y_v} = {p\over p+y_\mu}.
\label{thetaHB}
\end{equation}
By replacing the corresponding value of $y_\mu$, this relation yields the
value of $\theta$ for each specific transition.

\subsection{TSS at the low-density Mott transition in the
1D hard-core Bose-Hubbard model}
\label{tsshc1dbh}

We now show how TSS emerges at the low-density Mott transition
of the one-dimensional BH model, which is also of experimental relevance in
optical lattices, see, e.g., Refs.~\cite{BDZ-08,PWMMFCSHB-04,KWW-05,CFFFI-09}.
In particular, we consider its hard-core limit $U\to\infty$, which implies
that the particle number is restricted to the values $n_i=0,1$.  In this
limit, analytical results can be obtained by exploiting exact mappings into
the so-called XX chain model in the presence of a space-dependent external
field,
\begin{eqnarray}
H_{\rm XX} = - {1\over 2}\sum_i \left( \sigma^x_i \sigma^x_{i+1}
+ \sigma^y_i \sigma^y_{i+1} \right)
- \mu \sum_i \sigma^z_i  - \sum_i V(x_i) \sigma^z_i
\label{XXH}
\end{eqnarray}
(which is the Hamiltonian (\ref{Isc}) with $\gamma=0$).  The Pauli spin
matrices are related to the boson operators of the Bose-Hubbard Hamiltonian
$H_{\rm BH}$, cf.\ Eq.~(\ref{bhm}), by $\sigma^x_i = b_i^\dagger + b_i$,
$\sigma^y_i = i(b_i^\dagger - b_i)$, $\sigma^z_i = 1-2b_i^\dagger b_i$.
Actually we have that $H_{\rm XX}= 2 H_{\rm BH}(J=1,U\to\infty)$.  One can
then map it into a model of free spinless fermions by a Jordan-Wigner
transformation, given by Eq.~(\ref{sfi}) for $\gamma=0$, see, e.g.,
Ref.~\cite{Sachdev-book}.

In the absence of the trap, the 1D hard-core BH model has three phases: two
Mott insulator phases, for $\mu>1$ with $\langle n_i\rangle=0$ and for
$\mu<-1$ with $\langle n_i\rangle=1$, separated by a gapless superfluid phase
for $|\mu|<1$.  Therefore, there are two quantum transitions at $\mu=\pm 1$,
with $z=2$ and $y_\mu=1/\nu=2$. In the following we consider the low-density
Mott transition at $\mu=1$, for which we can
analytically show the existence of the TSS limit, and compute the
corresponding scaling functions. The results provide a further analytical
check of the TSS theory in quantum transitions, in a case with dynamic
exponent $z\ne 1$, whose quantum critical behavior is not described by a
$d+1$-dimensional conformal field theory.

The trap-size dependence, and corresponding TSS, turns out to be more subtle
at the $\langle n_i\rangle=1$ Mott insulator to superfluid transition, i.e.,
at $\mu=-1$.  This is essentially due to the fact that at $\mu=-1$ there is an
infinite number of level crossings as $l\to\infty$.~\footnote{For $\mu<1$, the
  ground state contains all the $\eta$-fermions diagonalizing the Hamiltonian
  with $\omega_k<0$, cf.\ Eq.\ (\ref{hdiagXX}).  In the presence of the
  trapping potential (\ref{potential}), level crossings of the lowest states
  occur in the $\mu$-$l$ plane separating the region with $N \equiv \sum_i
  \langle b_i^\dagger b_i \rangle =n$ from $ N =n+1$.  Since, in the absence
  of the trap potential and for $\mu<1$, the ground state has a finite {\em
    density\/} $N/L>0$, for fixed $\mu<1$ the lowest states 
  show an infinite number of level crossings as $l\to\infty$ (after
  $L\to\infty$) where the gap vanishes.}  Results will be presented 
elsewhere.

In the fermion representation the Hamiltonian can be easily diagonalized:
introducing new canonical fermionic variables $\eta_k=\sum_i \varphi_{ki}
c_i$, $k=0$, $1$, $\dots$, where $\varphi_{ki}$ satisfies
\begin{equation}
A_{ij}\varphi_{kj} = \omega_k \varphi_{kj}
\label{xxeq}
\end{equation}
and $A_{ij}$ is the matrix defined in Eq.~(\ref{sfi}), 
we obtain 
\begin{equation}
H = \sum_k \omega_k \eta_k^\dagger \eta_k.  
\label{hdiagXX}
\end{equation}
The ground state
contains all $\eta$-fermions with $\omega_k<0$, therefore the gap is
\begin{equation}
\Delta= \min_k |\omega_k|.
\label{deltaBH}
\end{equation}
A nontrivial TSS limit around $\bar{\mu}\equiv\mu-1=0$, i.e., at the
transition between a low-density superfluid and the empty vacuum state (named
$\langle n_i\rangle=0$ Mott phase above), is obtained (for small $|\omega_k|$)
by rescaling
\begin{equation}
x = l^{p/(2+p)} X, \quad \bar{\mu}
= l^{-2p/(2+p)} \mu_r, \quad \omega_k = l^{-2p/(2+p)} \Omega_k.
\label{rescBH}
\end{equation}
Indeed, neglecting terms which are suppressed in the large-$l$ limit,
Eq.~(\ref{xxeq}) becomes
\begin{equation}
\left(2 X^p - \partial_X^2\right)\varphi_k(X) =
(\Omega_k - 2\mu_r) \varphi_k(X).
\label{trapscaleqxx}
\end{equation}
This shows that $\theta=p/(2+p)$, in agreement with Eq.~(\ref{thetaHB}).

Moreover, this implies that any energy scale, and in particular the gap
$\Delta=E_1-E_0$, must behave as
\begin{equation}
\Delta \approx l^{-2\theta} {\cal D}(\mu_r),
\qquad \mu_r=l^{2\theta} \bar{\mu},
\label{deltaxxmu1}
\end{equation}
which agrees with the RG scaling equation (\ref{Deltasca}), since $z=2$ and
$\nu=1/2$. For $p=2$, by solving Eq.~(\ref{trapscaleqxx}), we obtain
\begin{equation}
{\cal D}(\mu_r)= \min_k| 2^{3/2} (k + 1/2) + 2\mu_r|,\quad
k=0,1,\dots;
\label{dmur}
\end{equation}
${\cal D}(\mu_r)$ is a triangle wave for $\mu_r\le0$ and it is linear for
$\mu_r\ge -1/\sqrt{2}$.

\begin{figure}[tbp]
\begin{center}
\leavevmode
\includegraphics*[scale=\graphicscale]{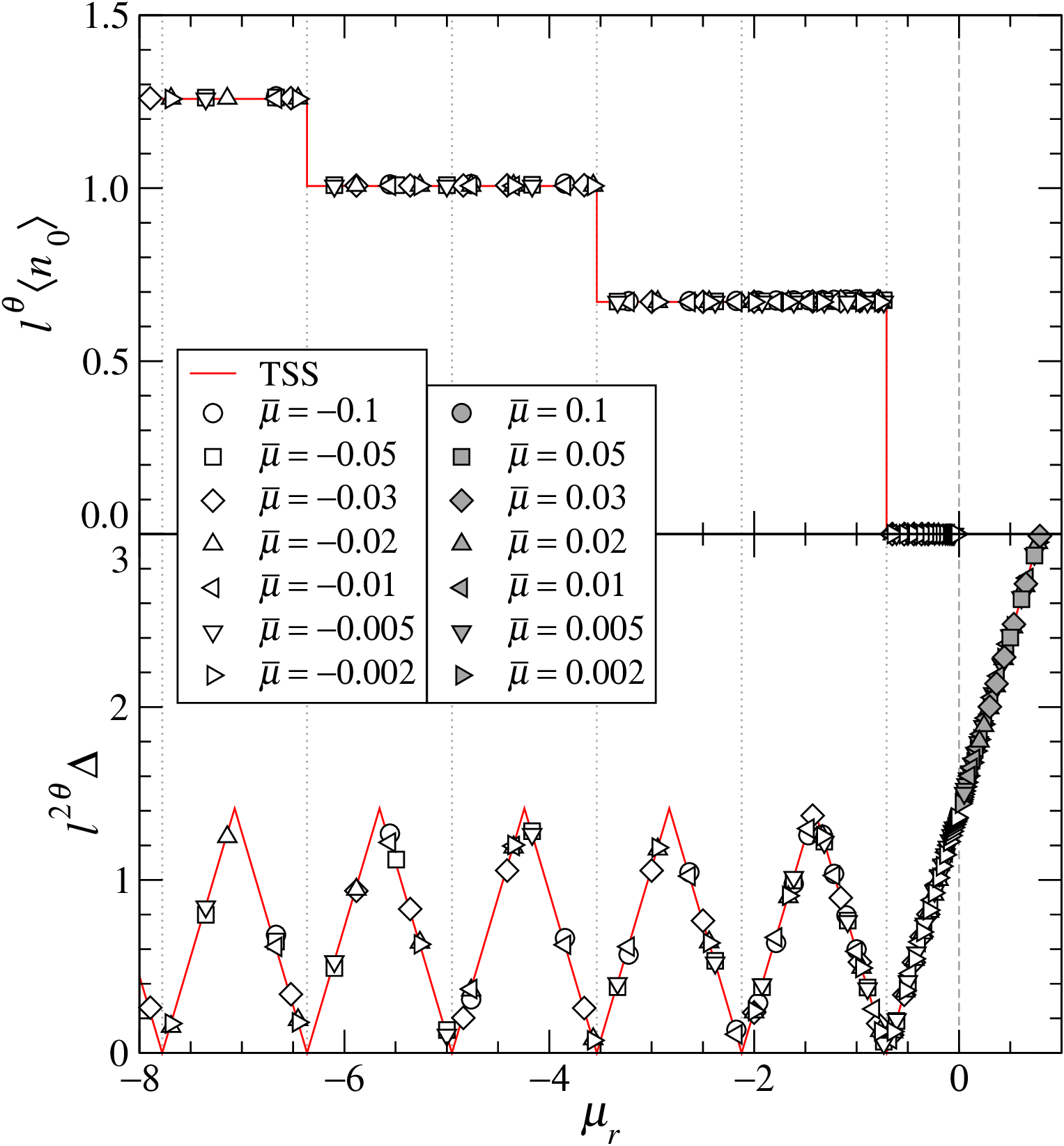}
\caption{(Color online)
  TSS of the gap (below) and the particle density in the middle of the
  trap (above) for $p=2$ and several values of $l\ge10$ and $\bar\mu$.
  The   lines correspond to Eqs.~(\ref{dmur}) and (\ref{n0xx}).
  Scaling corrections turn out to be very small. }
\label{XXfig}
\end{center}
\end{figure}

The TSS of the particle density in the middle of the trap is obtained by
computing
\begin{equation}
\langle n_0 \rangle = l^{-\theta} \sum_k |\varphi_k(0)|^2 \langle
\eta^\dagger_k \eta_k\rangle,
\label{n0exp}
\end{equation}
where $\varphi_k(X)$ are the normalized
eigenfunctions of Eq.~(\ref{trapscaleqxx}); $\langle \eta^\dagger_k
\eta_k\rangle=1$ if $\Omega_k<0$ and 0 otherwise; since
$\varphi_k(X)=(-1)^k\varphi_k(-X)$, only even $k$s contribute.  For $p=2$ we
obtain the sum
\begin{equation}
l^\theta \langle n_0\rangle \equiv
(2^{1/4}/\sqrt{\pi}) \sum {[(2j-1)!!]^2/(2j)!}
\label{n0xx}
\end{equation}
over integer $j\ge 0$ satisfying $2^{1/2}(2j+1/2) + \mu_r <0$.  Again, this
result agrees with the TSS theory, taking into account that the RG dimension
of the particle density is $y_n=d+z-y_\mu=1$.  Results from numerical
diagonalization at $p=2$ are shown in Fig.~\ref{XXfig}; they fully support the
above TSS behaviors.  Note the peculiar plateaux and the discontinuities in
the particle density at negative values of the scaling variable $\mu_r\equiv
l^{2\theta} \bar{\mu}$.  For $\mu_r\to-\infty$, $\langle n_0\rangle \approx
\sqrt{|2\bar\mu|}/\pi$, which matches the critical behavior for $\bar{\mu}<0$
in the absence of the trap~\cite{Sachdev-book}.

Numerical results for higher power laws of the potential turn out to be in
full agreement with the predictions of TSS and are qualitatively similar to
the results for $p=2$, including those for $p\to\infty$ which effectively
corresponds to the finite-size scaling of the homogeneous system without
confining potential.

\section{Conclusions}
\label{conclusions}

We have developed a trap-size scaling (TSS) theory for confined
particle systems at quantum transitions. We propose an extension of the
scaling laws of the homogeneous system to allow for the confining potential in
the large trap-size limit.  The quantum critical behavior of the confined
system is cast in the form of a TSS, resembling finite-size scaling
theory~\cite{FB-72,PV-02,GKMD-08}, with a nontrivial trap critical exponent
$\theta$, which describes how the length scale $\xi$ of the critical modes
diverges with increasing trap size, i.e., $\xi\sim l^{\theta}$,
at the quantum critical point. This provides
a theoretical framework to describe the quantum critical behavior of confined
systems.

We have shown by explicit analytical computation how TSS emerges in
the quantum XY chain, where an inhomogeneity analogous to the one
arising from a trapping potential in particle systems can be achieved
by considering a space-dependent transverse field.

Moreover, we have discussed this issue within the Bose-Hubbard model with a
confining potential, which is relevant for the description of cold atomic
gases in optical lattices.  In particular, we have presented some analytical
results for the low-density Mott transition in the hard-core limit of the
one-dimensional BH model. The results again support TSS, although the
corresponding TSS functions show a peculiar behavior, like discontinuities in
the scaling particle density, which are clearly related to the quantum nature
of the transition.

\medskip
Helpful discussions with P. Calabrese are gratefully acknowledged.

\end{document}